# Virtual Reality and Artificial Intelligence as Psychological Countermeasures in Space and Other Isolated and Confined Environments: A Scoping Review

**Author Information**
Name: Jennifer Sharp*
Affiliation: School of Psychology, Charles Sturt University, Bathurst, NSW, Australia.
ORCID: https://orcid.org/0009-0001-2305-1244
Email: jesharp@csu.edu.au

Name: Joshua Kelson
Affiliation: School of Psychology, Charles Sturt University, Bathurst, NSW, Australia. Email: jkelson@csu.edu.au
ORCID: https://orcid.org/0009-0000-7066-4206

Name: Daryl South
Affiliation: School of Education, Charles Sturt University, Wagga Wagga, NSW, Australia.
Email: dsouth@csu.edu.au
ORCID: https://orcid.org/0000-0002-9814-957X

Name: Anthony Saliba
Affiliation: School of Psychology, Charles Sturt University, Wagga Wagga, NSW, Australia.
Email: asaliba@csu.edu.au
ORCID: https://orcid.org/0000-0002-4823-4602

Name: Muhammad Ashad Kabir*
Affiliation: School of Computing, Mathematics and Engineering, Charles Sturt University, Bathurst, NSW, Australia. Email: akabir@csu.edu.au
ORCID: https://orcid.org/0000-0002-6798-6535

* Corresponding authors.

**Abstract**

Spaceflight is an isolated and confined environment (ICE) that exposes astronauts to psychological hazards, such as stress, danger, and monotony. Virtual reality (VR) and artificial intelligence (AI) technologies can serve as psychological countermeasures as they can digitally simulate immersive environments, interactive companions, and therapeutic experiences. Our study employs a scoping literature review approach to identify what is currently known about the use and effectiveness of VR and AI-based interventions as psychological countermeasures to improve mood or emotional states in adults in space or other ICEs. Additionally, this review aimed to identify gaps in the knowledge base and whether a systematic review with meta-analysis was warranted. The review included studies where the intervention was used or intended for use in space or other extraterrestrial environments (ICE). Our search strategy yielded 19 studies from 3390 records across seven major databases. All studies focused on VR-based interventions, with no eligible AI-based intervention studies found. VR interventions were found to be effective for relaxation and improving mood, emergency training, as an interactive communication platform, for comparing interior designs, and for enhancing exercise. There were improvements for measures of mood and emotion\n (e.g., anxiety and stress); however, user preferences varied, and some instances of cybersickness were reported. A systematic review with meta-analysis is not recommended due to the heterogeneity of results. There is significant scope for further research into the use of VR for a wider range of mood and emotion variables using standardised assessment instruments. Additionally, the potential application of AI as a psychological countermeasure warrants further investigation.

**Keywords**

Virtual Reality; Artificial Intelligence; Mental Health; Space; Isolated and Confined Environments; Wellbeing.

## 1. Introduction

There has been renewed interest in the space sector with the development of private space companies such as Space X, Virgin Galactic, and Blue Origin, and there are plans within the foreseeable future to extend humanity's presence in space with an orbital lunar gateway [1], and manned exploration of Mars [2]. As the population living and working in space grows, it is important to ensure that these environments are designed to maintain optimal psychological health.

The prevailing view of astronauts has historically been of an elite handful of exceptional individuals with 'the right stuff'; a rare combination of skill, bravery, luck, daring, and fearlessness [3]. Despite the public perception of astronauts as immune to the stresses and worries of the average person, mental health challenges in spaceflight were recognised in the early days of humanity's foray into space [4-6]. The psychological hazards of spaceflight and the resulting need for mitigation of these hazards are recognised by NASAs Human Research Roadmap as an essential issue to address for deep space travel [7-9].

Stressors, which are environmental factors that usually impact people negatively, are psychological hazards [10]. Figure 1 outlines some of the stressors encountered in the space environment.

**Figure 1.** *Stressors [10]*

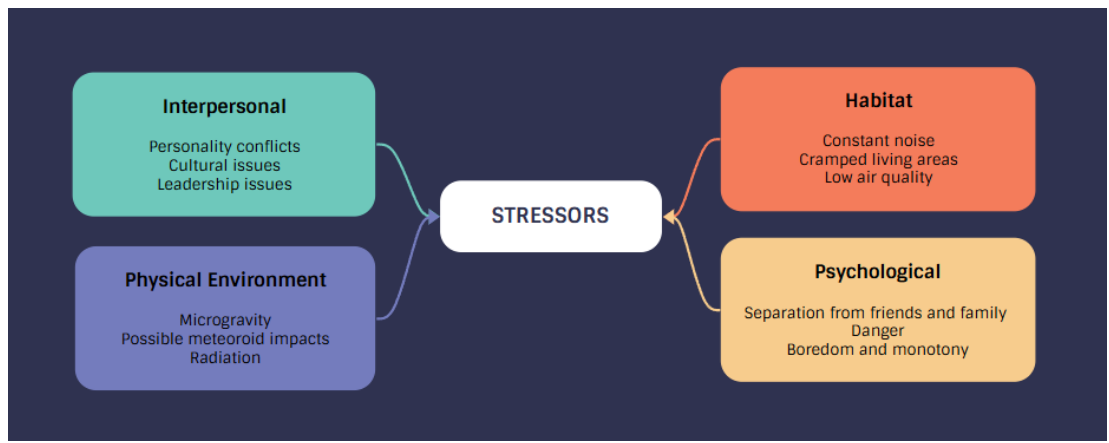

Strategies to combat these stressors are known as psychological countermeasures and are defined as "all actions and measures that alleviate the effects of the extreme living and working conditions of space flight on crew performance and behavior" [11]. Some examples of psychological countermeasures are included in Table 1.

**Table 1.** *Psychological Countermeasures [10]*

| Domains | Psychological Countermeasures |
|---|---|
| Psychological | Crew selection |
| | Coping skills training |
| | Self-monitoring, self-care |
| | Ongoing psychological monitoring and support, counselling |
| | Surprise gifts, entertainment, leisure time, celebrations |
| | Contact with family, friends, celebrities |
| | Support for family at home |
| Interpersonal | Pre-flight training: |
| | • group dynamics, crew-ground relationships, team effectiveness, cultural issues, languages, communication, social support, interpersonal skills, crew co-ordination |
| | Regular facilitated debriefs to resolve issues |

Approaches planning for multiple countermeasures are likely to be most effective, as despite individual screening and selection, psychiatric issues such as mood, stress, and sleep disorders have been documented in people in both space and Antarctica [12].

Research into the space environment is limited by high costs and restricted access, and researchers must look beyond space for knowledge about how humans can survive and thrive in isolated and confined environments (ICEs). Space analogues, which are naturally occurring or laboratory-controlled terrestrial environments that share varying degrees of similarity to the space environment, can provide valuable information about how individuals and groups function in isolation and confinement [13]. Analogues can include expeditions to geographically remote locations, isolation chambers, submersible habitats, habitats in extreme environments, and polar stations [13], as well as ocean-based oil-drilling platforms, prisons, and caves [10]. Some of the shared psychological and social characteristics of space and analogue ICEs include fatigue, stress, fear, crowding, and isolation from family members and the individual's cultural group [14].

Technology plays a fundamental role in completing mission tasks and providing life support in both space and ICE's [15]. In addition to providing the environmental conditions to sustain life and enable individuals to carry out their work, technology can also be used to meet crewmember's psychological needs. Virtual reality (VR) and artificial intelligence (AI) hold appeal for use as potential psychological countermeasures due to the rapid advances that have been made in these technologies over recent years, as well as their versatility of use for both training and entertainment [16]. VR consists of a head-mounted display that provides a 360° visual field and auditory output, as well as handheld devices that provide haptic feedback and allow users to control their movements within the VR environment, resulting in a 3-dimensional, realistic experience for the user [17]. VR provides access to experiences not readily available in a confined environment, such as immersive simulations, games, concerts [16], and opportunities for shared experiences with children, friends, and family members. Gushin and Ryumin [18] suggest the use of VR as a psychological adjunct to improve sensory deprivation and monotony in space through exposure to views of Earth regions, leisure activities, and the creation of a virtual environment to provide a sense of personal space. VR has also been investigated as a potential countermeasure for anticipated stressors in space, through graded exposure to smoke in the International Space Station [19] and using nature scenes for relaxation [20].

AI is another rapidly advancing technology that holds promise as a potential psychological countermeasure. AI is defined as a system which has the ability to interpret external data, learn from that data, and to flexibly adapt what it has learned to achieve specific goals [21]. Social robots are artificial agents designed to interact with people and elicit social responses from them [22], and a social robot utilising artificial intelligence has been trialed in space [23]. Gushin and Ryumin [18] proposed the use of voice-interfaced social robots or virtual assistants to address loneliness and social isolation in space, where in addition to interacting with equipment, they could use speech and facial analysis to monitor emotional wellbeing, present news and entertainment, engage in voice-interactive games, and use active listening techniques to provide psychological support to astronauts. A mission to Mars is anticipated to take up to 3 years [24], and most interactions with people on Earth during this time are likely to be impacted by latency, which is the delay between sending and receiving communications [25]. Gushin and Ryumin [18] noted that interactions with a social robot or virtual agent, which would be instantaneous, would be of particular use when interaction with family and friends was impacted by communication delays.

Given the speed at which VR and AI technologies have advanced over the past decade and the relative novelty of their use as a psychological countermeasure with this population, a scoping review was the most appropriate approach to identify broadly what is currently known about this topic, and to identify gaps in the knowledge and examine how research in this field is being conducted [26]. No current or proposed scoping reviews on this topic were identified from preliminary searches. Smith [27] conducted a scoping review to map the published literature within the field of space psychology, and Gatti and Palumbo [28] conducted a review of countermeasures for long-duration space exploration. A qualitative literature review identified the potential of VR as a psychological countermeasure for the space environment [29]; however, our scoping review focuses on the use of VR and AI as psychological countermeasures in space or another ICE. There are a wide range of multidisciplinary AI and VR applications within the literature; however, the scope of this review was restricted to psychological countermeasures impacting mood and emotion as these factors are critical to astronaut well-being and mission performance in ICE.

**Review Questions**

The primary research question for this scoping review was: '*What is currently known about the use and effectiveness of VR and AI to improve mood or emotional states in people in space or another isolated and confined environment?*'. The secondary research question was '*What research gaps are there in the knowledge base, and is a systematic review with meta-analysis warranted?*'

2. **Method**

This review was conducted in accordance with the scoping review methodology published by the Joanna Briggs Institute (JBI) [30], and the Preferred Reporting Items for Systematic Reviews and Meta-Analyses checklist adapted for Scoping Reviews (PRISMA-ScR) was followed [31]. The protocol for this scoping review was registered with the Open Science Framework on 2nd July 2024, and can be accessed here: https://osf.io/zj34e/?view_only=8a89958566464991a885884327bc22b2.

**2.1 Eligibility Criteria**

The population eligible for inclusion in this scoping review were adults of any nationality who were astronauts, or who were living and working in an ICE, such as an space simulation or analogue environment. Within this population, only studies using VR or AI-based interventions as psychological countermeasures to improve mood or emotional states were included. Studies with participants who were not astronauts or living and working in an ICE were included when the purpose of the research was to assess the effectiveness of an intervention specifically for use in space or another ICE. Technology included all head-mounted displays or Cave Automatic Virtual Environments (CAVE) VR systems, and all AI-based interventions where a mood or emotional state outcome was measured. Studies were included if they used any type of comparator, or any standardised or unstandardised measures of mental health outcomes, and any acceptability and useability measures of the technologies. All qualitative, quantitative, and mixed-methods research designs were included. The quality of the methodology did not exclude any studies. All of the studies included were primary studies, published as conference proceedings or peer-reviewed journal articles, and written in English. No other grey literature was included.

**2.2 Search Strategy**

The following scientific research article databases were searched up until April 2024: Cochrane Library, IEEE Explore, PsycINFO, PubMED, Scopus, and Web of Science. The first 10 pages of Google Scholar were also searched [32, 33]. The keywords in Table 2 were searched, with the term from each of the three categories being paired with the term from every other category. Articles with these terms in the title or abstract were searched. All of the keywords and terms were adapted for each database searched, and the reference lists of included studies were searched for additional relevant literature.

**Table 2.** *Search terms*.

| Cochrane Library Search |
|---|
| *Technology Search Terms* |
| 1. Virtual reality |
| 2. Artificial intelligence |
| AND *Mood or Emotion Search Terms* |
| 3. Psych* |
| 4. Countermeasure* |
| 5. Mental |
| 6. Mood |
| 7. Emotion |
| 8. Social |
| 9. Human |
| 10. Conflict |
| 11. Stress |
| AND *Demographic Of Interest Search Terms* |
| 12. Isolat* |
| 13. Spaceflight |
| 14. Astronaut |

*Note*. * Indicates multiple permutations of the word searched, e.g., psychology, psychological.

## 2.3 Article Selection

Only published studies were included. Duplicated results were removed from the data set for all databases, and the titles and abstracts of the remaining records were reviewed. Using the eligibility criteria, appraisals of the full text and agreement on final inclusion were conducted by two researchers. Differing views on inclusion were resolved through discussion until a consensus was reached.

## 2.4 Data Extraction

Reviewer One (JS) extracted data from eligible studies into a standardised coding sheet, and data were then checked by Reviewer Two (DS). The extracted data items included the following data types:

1. Study source: article title, year published, and author surnames.

2. Study design: methods used, comparison types, and any measured time-points (e.g., pre, during, and post-test) of outcome variables.
3. Population of interest: number of participants, age, gender, and nationality.
4. Intervention details: type of VR or AI intervention, type of equipment used (hardware and software), protocol used, length of intervention, frequency and duration of intervention sessions, and whether therapist or self-administered.
5. Effectiveness outcomes: names of measures used, outcomes, and effect sizes.
6. User experience outcomes: measures and ratings of usability, safety of intervention, rate of attrition, acceptability of use, and likelihood of future use.

For interventions with multiple components or exposures, the number of participants who started and then discontinued the use of the AI or VR intervention, or who completed initial but not subsequent measures, is defined as attrition in this review.

**2.5 Quality Assessment**

This scoping review evaluated studies with a diverse range of methodologies. To guide future research on this topic and to meet the aims of the review, the Mixed Methods Appraisal Tool (MMAT) [34] was used by two reviewers (JS, AK) to appraise the quality of the included studies. The MMAT is a questionnaire tool designed to guide the user in critically assessing the methodological quality of each study based on five design types: qualitative, quantitative descriptive, randomised control, non-randomised, and mixed methods [34].

**2.6 Data Analysis**

Given the heterogeneity observed in data and study designs across the included articles, this scoping review used a qualitative narrative synthesis approach to summarise the results and answer the research questions.

## 3. Results

### 3.1 Study Selection

See Figure 2 for the study selection process flowchart, and Table 3 for a summary of the included studies, all of which were VR-based.

**Figure 2.** *Study Selection Process*

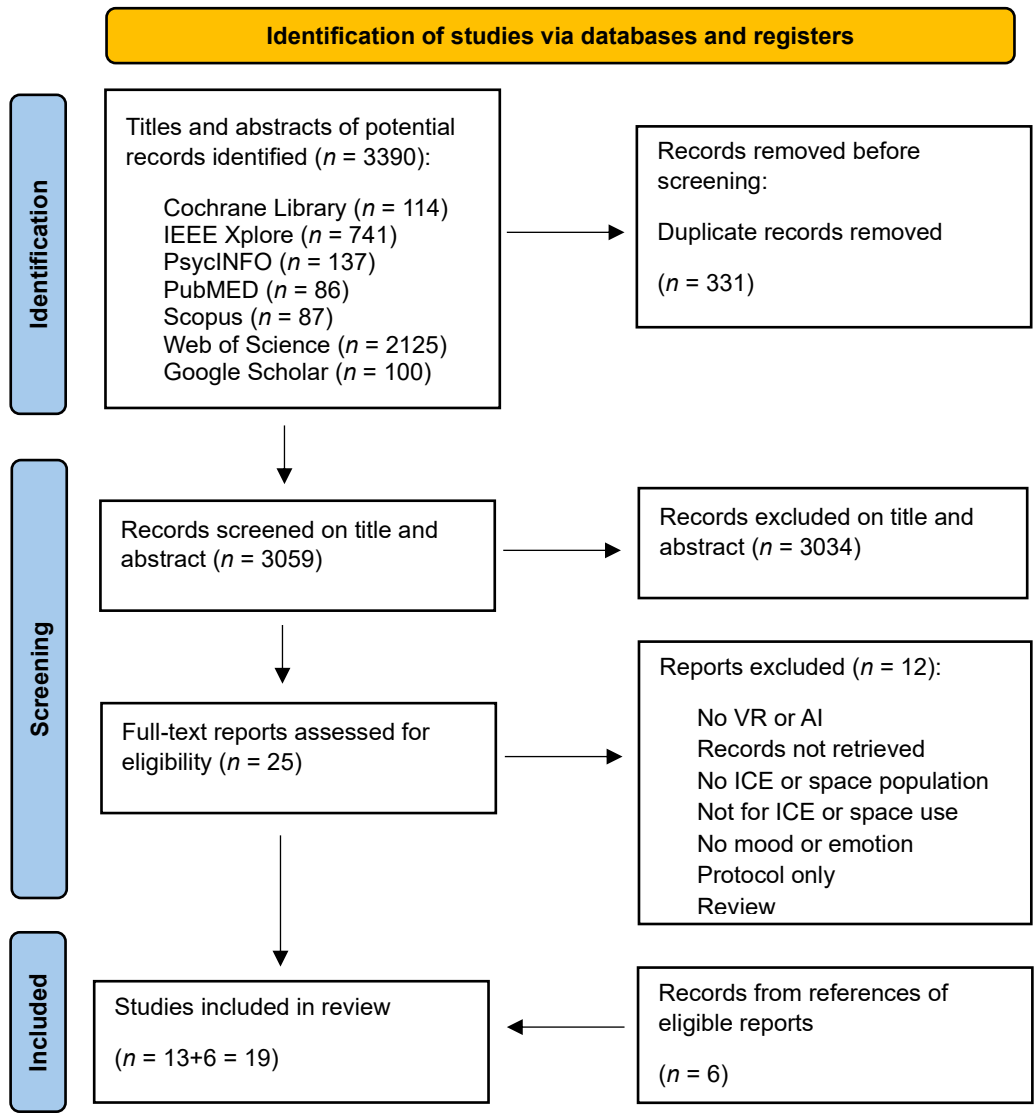

**Table 3.** *Studies Using Virtual Reality to Impact Mood or Emotion in Space or Another Isolated and Confined Environment (ICE)*

| Study | Number of participants (% Female) | Study Location | Mean Age in Years (SD) | Measures | Intervention | Effectiveness |
|---|---|---|---|---|---|---|
| Abbott & Diaz-Artiles (2022) | 10 (30) | Lab | 23.1 (2.8) | Mood - Positive And Negative Affect Scale (PANAS) Anxiety - State-Trait Anxiety Inventory 6-item (STAI-6) | Subjects completed a stress task then navigated a 15 minute VR nature environment with scents followed by the same procedure with or without scents (control). | Reduced negative affect ($p = 0.003$) and anxiety ($p = 0.001$) scores after the scented VR condition. |
| Anderson et al. (2017) | 18 (50) | Lab | 32 (12) | Stress – Electrodermal activity (EDA) Heart Rate Variability (HRV) --- Mood - Positive And Negative Affect Scale (PANAS) VR Presence - Modified Reality Judgment and Presence Questionnaire (MRJPQ) VR Value - Value of VR Questionnaire (VVR) | Subjects completed a stress task and then viewed three, 15 minute VR nature scenes (beaches, Irish countryside, indoor control). | Natural scenes reduced stress ($p = 0.03$), and negative affect ($p = 0.005$, and $p = 0.03$) and people found VR relaxing ($p = 0.015$). No significant differences for HRV. |
| Anderson et al. (2023) | 43 (16) | Simulation and Analogue | Not specified | Mood - Positive And Negative Affect Scale (PANAS) VR Presence - Modified Reality Judgment and Presence Questionnaire (MRJPQ)VR Value - Value of VR Questionnaire (VVR) Interviews Restorativeness - Perceived Restorativeness Scale (PRS-11) Mood - Profile of Mood States (POMS) | Military base: subjects viewed one of several, 15 – 20 minute VR nature scenes each week (minimum). Simulation 1: subjects viewed two VR natures scenes in counterbalanced order three weeks apart, then viewed additional VR scenes. Simulation 2: subjects viewed 16, 15 - 20 minute VR natures scenes. | Nature scenes were more restorative ($p = 0.010$). No significant difference in affect between scenes, mood, or value ratings. Interviews were previously reported (Lyons et al., 2020). |

| Study | Number of participants (% Female) | Study Location | Mean Age in Years (SD) | Measures | Intervention | Effectiveness |
|---|---|---|---|---|---|---|
| Finseth et al. (2018) | 20 (0) | Lab | 22.5 (2.2) | Stress – Heart Rate Variability (HRV) Blood Pressure (BP) --- Cognitive workload - NASA Task Load Index (NASA -TLX) Stress - Short Stress State Questionnaire (STSQ) Time-to-complete | Subjects responded to an emergency VR scenario of locating and putting out a fire with heavy smoke in the ISS, following practice with light smoke or no smoke VR scenarios. | Prior exposure to a mild stressor reduced stress responses to a more severe stressor. HRV measures showed lower autonomic nervous system activation for the treatment group. Task engagement was lower for light smoke than no smoke conditions (p = .041). Testing had greater temporal (p = .02) and physical (p = .048) demands than training. VR training group took longer to complete the emergency condition (p = .05, d = 1.19). No other significant differences. |
| Finseth et al. (2020) | 62 (24.2) | Lab | 20.6 (2.6) | Stress - Post Task Stress Reaction (PSTR) Mood - Positive And Negative Affect Scale (PANAS) Stress - Short Stress State Questionnaire (SSSQ) Cognitive workload - NASA Task Load Index (NASA -TLX) Stress - Free Stress Scale of Events (FSSE) | Subjects had a familiarisation session with the VR ISS environment, then three exposure trials (low, medium, high stress) to an emergency VR scenario locating and putting out a fire on the ISS. | The stress training conditions were significantly differentiated between low, medium, and high for subjective stress (p < 0.001, d = 1.58, p < 0.001, d = 3.02), distress between high and medium levels (p = 0.002, d = 1.17), and subjective workload between high and medium levels (p < 0.001, d = 1.18). Mood was not significantly different. |
| Finseth et al. (2021) | 57 (36) | Lab | 28.7 (4.6) | Stress – Heart Rate (HR) Heart Rate Variability (HRV) --- Distance to fire | Subjects completed three calibration trials of a VR emergency scenario smoke in the ISS, then five stress induction trials to compare skill only, graduated, and adaptive training conditions. | Stress in all conditions was reduced and was most reduced in the adaptive condition (HRV, LF/HF p = .032, d = 0.25). Heart rate was lower at the end of the trials for the graduated (p = .006, d = 0.55) and adaptive (p = .014, d = 0.49) conditions. Distance to fire was not significant. |
| Finseth et al. (2022) | 40 (7.5) | Lab | 20.5 (2.6) | Stress – Heart Rate (HR) | Subjects completed a VR tutorial about the ISS and | Manipulating stress levels was effective. Higher stress conditions had |

| Study | Number of participants (% Female) | Study Location | Mean Age in Years (SD) | Measures | Intervention | Effectiveness |
|---|---|---|---|---|---|---|
| | | | | Heart Rate Variability (HRV) Blood Pressure (BP) Electrodermal activity (EDA) ---Cognitive workload - NASA Task Load Index (NASA -TLX) Stress - Short Stress State Questionnaire (SSSQ) Stress - Free Stress Comparison (FSC) Stress - Post Task Stress Reaction (PTSR) Anxiety - State-Trait Anxiety Inventory (STAI) | emergency fire procedures, then practiced the procedures, then completed low, medium, and high stress VR conditions. | higher anxiety (p = 0.009, d = 0.98), subjective stress (p < 0.001, d = 1.71, p < 0.001, d = 2.91), and subjective workload (p < 0.001, d = 1.15), as well as physiological measures of stress (EDA, p = 0.002, d = 0.93, HR p = 0.017, d = 0.90, HRV RMSSD p = 0.04, d = 0.48). |
| Finseth et al. (2023) | 41 (17) | Lab | 20.9 (6.5) | Stress - Free Stress Scale (FSS) Physiological personalised stress model: Electrocardiogram (ECG), Respiration (RSP), Electrodermal Activity (EDA), Noninvasive Blood Pressure (NIBP) | Subjects completed a 20 minute VR training session, then three trials for low, medium, and high stress VR emergency fire scenarios, or a non-VR condition consisting of a 3 minute tutorial and low, medium, high stress trials. | For the VR condition, stressor levels were differentiated by low, medium, and high (p <.001, d = 3.02), and the physiological model had better stress prediction than the generalized model (82% vs 62%). |
| Finseth et al. (2024) | 65 (44.6) | Lab | 28.7 (4.6) | Stress – Heart Rate (HR) Heart Rate Variability (HRV) Blood Pressure (BP) Electrodermal activity (EDA) --- Stress - Post Stress Task Reaction Measure (PSTRM) Stress - Short Stress State Questionnaire (SSSQ) | There were three groups; skill only, gradual stress increase, or adaptive personalised stress increase. Subjects completed three calibration trials then five experimental condition trials where subjects responded to a fire emergency on the ISS. | Trials differentiated low, medium, and high stress levels (p < .001, d = 2.27). Stress was reduced in all conditions (HRV PNN50, RMSSD all significant or marginally significant), and marginally lower in the adaptive condition (LF/HF). Task engagement was higher in the adaptive VR condition (p=.021, d=0.53), worry was lower in all VR conditions (all p<.05), and anxiety ratings were lower in the adaptive |

| Study | Number of participants (% Female) | Study Location | Mean Age in Years (SD) | Measures | Intervention | Effectiveness |
|---|---|---|---|---|---|---|
| | | | | Anxiety - State-Trait Anxiety Inventory (STAI) Cognitive workload - NASA Task Load Index (NASA -TLX) | | and graduated VR conditions (p=.005, d=0.68, p=.023, d=0.53). Lower HR in adaptive VR condition (HR, p < .001, d = 0.93). Preliminary results presented in Finseth et al. (2021). |
| Firth & Jayadas (2022) | 6 (66) | Lab | 23.33 (2.66) | Stress – Blood Pressure (BP) Heart Rate (HR) --- Anxiety - Modified State-Trait Anxiety Inventory (mSTAI) Overall perceptions and impressions of the space. | Subjects completed a VR practice session, then viewed two VR environments (biophilic and non-biophilic) for 5 to 10 minutes each in counterbalanced order while exploring the environments and verbalizing their thoughts. | Positive perceptions were significantly higher and negative perceptions were significantly lower for 5 items in the biophilic environment (p< .05). No significant physiological differences. Impressions - more negative perceptions expressed in the non-biophilic VR environment. |
| Galunder et al. (2018) | 6 (Not specified) | Simulation | Not specified | Social connection - Circles of Closeness (COC) Team dynamics - Mentalblock game Cognitive workload - NASA Task Load Index (NASA -TLX) Cognitive workload - Bedford Scales | Subjects interacted with each other in a VR environment three to four times per week for 12 months. | Interacting via VR increased the ratio of positive vs negative language used (p<0.001). Some scale values not reported. |
| Keller et al. (2022) | 5 (0) | Lab | 35 (9) | VR Value - Value of VR Questionnaire (VVR) Emotional Distress - General Health Questionnaire (GHQ) Stress - Perceived Stress Scale (PSS) Motivation - Sport Motivation Scale-6 (SMS-6) Anxiety - State-Trait Anxiety Inventory (STAI) Feelings - Feeling Scale (FS) Feelings - Felt Arousal | Subjects completed 30 minute rowing tasks with or without VR six days per week for two weeks, then had a one month break, then 6 days per week for two weeks in the opposite condition. | Negative affect in the non-VR group increased over time (p = 0.009) and negative affect (p = 0.026) and fatigue (p = 0.015) were higher in the non-VR group in a training condition. VR was more restorative in 2 subscales (both p<0.001). No significant differences in other relevant measures. |

| Study | Number of participants (% Female) | Study Location | Mean Age in Years (SD) | Measures | Intervention | Effectiveness |
|---|---|---|---|---|---|---|
| | | | | Scale (FAS) Effort - Rating of Perceived Exertion (RPE) Exercise Affect - Physical Activity Affect Scale (PAAS) Restorativeness - Perceived Restorativeness Scale (PRS-11) VR Presence - Spatial Presence Experience Scale (SPES) | | |
| Lyons et al. (2020) | 18 (44.4) | Simulation | Not specified | Semi-structured interviews. | Subjects viewed each VR nature scene at least once, then at will during months four to eight of the simulation missions. | This study used a qualitative analysis. VR was generally perceived positively. Precedes Anderson et al. (2023). |
| Ott et al. (2016) | 6 (50) | Simulation | Not specified | Social connection - Circles of Closeness (COC)Stress - Perceived Stress Questionnaire (PSQ). Useability - System Usability Scale (SUS) Presence, Task Difficulty, and Discomfort - Short Feedback Questionnaire (SFQ) | Subjects interacted with family and friends through a VR platform. Each session lasted 20-30 minutes. | VR group had greater feelings of closeness (p<.05) and satisfaction (p<.05) with friends and family. Stress decreased marginally. |
| Pochwatko et al. (2023) | 5 (40) | Simulation | Not specified | Stress – Heart Rate (HR) Electrodermal activity (EDA) --- Wellbeing - Short Mental Wellbeing Scale (SMWS) Emotions - The Pictorial Self-Assessment Mannequin (SAM) Response to VR - Multi-dimensional feedback tool | Subjects participated in each of the cinematic, interactive, and nature VR experiences on day eight, 10, and 12 of the simulation. The cinematic and nature VR scenes were seven minutes in length each. | Wellbeing changes were marginal, mixed physiological findings from 2 participants (EDA). Artistic VR reduced asthenia symptoms more than nature VR, interactive VR most effective. |

| Study | Number of participants (% Female) | Study Location | Mean Age in Years (SD) | Measures | Intervention | Effectiveness |
|---|---|---|---|---|---|---|
| | | | | Presence - Short Presence Scale (SPS) Asthenia symptoms Interviews | | |
| Stankovic et al. (2019) | 6 (Not specified) | Simulation | Not specified | VR Presence - Modified Reality Judgment and Presence Questionnaire (MRJPQ) Restorativeness - Perceived Restorativeness Scale (PRS) Mood - Profile of Mood States (POMS) | Subjects viewed 16 VR nature scenes at least once each in the order and time of their choice. | Filmed natural scenes were more immersive and restorative than rendered scenes. No significance tests reported. |
| Winn et al. (2023) | 40 (72.5) | Lab | Median: 25.5 Range: 18-57 | Emotion – Heart Rate (HR) Pupil dilation --- Anxiety - Modified State-Trait Anxiety Inventory (mSTAI) Cognitive workload - Modified NASA Task Load Index (mNASA-TLX) | Subjects completed a 5 minute practice session, then two VR environments (biophilic or non-biophilic, < 15 minutes each) in randomly presented order and were prompted to verbalise their thoughts about the environments. | People were calmer, more content in the biophilic VR (both $p < 0.0001$), and more nervous and indecisive in the non-biophilic VR (both $p < 0.0001$). Greater pupil dilation ($p < 0.0001$) in biophilic VR. Better performance in biophilic VR ($p = 0.006$) with less effort ($p = 0.02$). Higher intention to spend time in the biophilic VR ($p < 0.0001$) and higher satisfaction ($p < 0.0001$). No significant differences in heart rate or other measures. |
| Wu, Morie, et al. (2016) | 6 (Not specified) | Simulation | Not specified | Social connection - Circles of Closeness (COC) | Subjects communicated with family and friends through a VR platform. | Feelings of closeness and satisfaction were greater (both $p<.05$) when using VR to interact with family and friends than traditional non-VR methods. |
| Wu, Ott, et al. (2016) | 6 (Not specified) | Simulation | Not specified | Social connection - Circles of Closeness (COC) Social connection - General connectedness (GC) | Subjects used the VR platform to interact with family and friends three times per week for 45 minutes per session. | COC reported above. VR group felt closer to humanity ($p<.001$) and that time passed more quickly ($p<.01$) than non-VR group. Involvement with family and friends |

| Study | Number of participants (% Female) | Study Location | Mean Age in Years (SD) | Measures | Intervention | Effectiveness |
|---|---|---|---|---|---|---|
| | | | | Stress - Perceived Stress Questionnaire (PSQ) Time perception | | associated with lower stress (p<.01). |

*Note*: Simulation = a simulated space mission where participants lived in an isolated and controlled environment for a period of time. HRV measures: RMSSD = Root Mean Square of the Successive Differences. LF/HF = the ratio of Low Frequency to High Frequency. PNN50 = The proportion of NN50 divided by the number of (R-R) intervals

### 3.2 General findings

There were 19 studies that met the scoping review eligibility criteria. Despite preliminary searches identifying a handful of studies using AI to impact mood or emotion in the space and ICE domains, no AI studies met all the selection criteria in this review. An earlier published protocol for ANSIBLE (A Network of Social Interactions for Bilateral Life Enhancement) indicated the use of artificial intelligence in the form of a virtual assistant that used a sociolinguistics-based engine to engage in social behaviours and to monitor wellbeing through verbal and written communications within the virtual environment [35]. However, it is unclear how this virtual agent was used, as the articles and conference papers based on this research do not provide further details [36-39]. All of the studies (n=19) included in our review used VR, and not AI, as the psychological countermeasure tool. See the summary of studies in Table 3 for relevant effect size and probability values from the included studies that reported them.

The included studies were published between 2016 and 2024, with 12 of the studies (63%) published from 2020 onwards. All of the studies were from the USA, with the exception of Pochwatko and Kopec [40], where the country was not specified, and Anderson and Stankovic [41], who included participants from a Canadian military facility.

There was wide variation in the applications of VR to space. Several studies (n=6) specifically focussed on the use of VR to directly impact mood or emotion [20, 40-44], but for other studies (n=13), mood or emotion were part of a broader suite of measures to assess the effectiveness of VR as an emergency training tool [19, 45-49], an interactive communication platform [36-39], to compare interior designs [50, 51], or to enhance an exercise regime [52]. Table 4 lists the areas of VR research.

**Table 4.** *Areas Of Research Using VR To Impact Mood or Emotion in Space or Another ICE*

| Research Areas | Authors (Year) |
|---|---|
| Enhancing the mood effects of the virtual environment using scents | Abbott and Diaz-Artiles (2022) |
| Graduated stress exposure training to reduce stress in space emergencies | Finseth et al. (2018) |
| | Finseth et al. (2020) |
| | Finseth et al. (2021) |
| | Finseth et al. (2022) |
| | Finseth et al. (2023) |
| | Finseth et al. (2024) |
| Alleviation of asthenia symptoms | Pochwatko et al. (2023) |
| Using a communication platform (ANSIBLE) for people to interact and improve interpersonal relationships | Ott et al. (2016) |
| | Wu, Morie, et al. (2016) |
| | Wu, Ott, et al. (2016) |
| | Galunder et al. (2018) |
| Test biophilic design elements for space habitats | Firth and Jayadas (2022) |
| | Winn et al. (2023) |
| Exposure to nature scenes for relaxation | Anderson et al. (2017) |
| | Anderson et al. (2022) |
| | Lyons et al. (2020) |
| | Stankovic et al. (2019) |
| Improve mood and motivation during exercise | Keller et al. (2022) |

**3.3 Participant Characteristics**

Participants were mainly comprised of students and university staff or were from the general population (reported in Table 3). Few studies (n=5) used astronaut-like participants [36, 45], who were representative of an astronaut population on characteristics such as age, gender balance, Science Technology Engineering and Mathematics (STEM) educational backgrounds, and fitness levels [45, 47, 48, 51, 52]. Only two of the studies described the participant's ethnicity; 60% Caucasian, 15% African American, 15% Hispanic or Latino in Finseth and Keren [19], and 76% European American/White, 12% Asian or Asian American, and 7% Hispanic or Latino in Finseth and Dorneich [49]. Participant's previous experience with VR was mentioned in three studies [20, 44, 50]. Eight of the studies reported on participants in actual ICEs, such as a military facility in Canada [41] and space simulation mission lasting from 14 days in length [40] to 12 months in length [41]. There were no astronaut participants used in any of the studies, although three studies used feedback from NASA personnel and astronauts in the development of projects [39, 46 , 47].

### 3.4 Research Designs and Comparators

Consistent with a nascent field of research, the majority of studies (n=11) were small, pilot-type studies that used within-groups designs [20, 40-44, 46, 47, 50-52]. Five studies used between-group designs [19, 36-39], and others (n=3) used between-and-within group designs [45, 48, 49]. For between-group designs studies (n=5), the comparators were unexposed control groups [19, 36-39].

Studies used a mixture of objective, physiological measures of stress, and subjective, self-report measures of mood and emotion (see Table 3). Physiological measures were recorded continuously during testing sessions [19, 20, 40, 45, 47-49, 51] or in pre-test and pos-test comparisons [50]. There was a range of self-report measures of mood and emotion used (see Table 3), and these were variously administered a pre-test and post-test exposure to condition [19, 20, 41, 42, 44-47, 52], during the testing session [36, 51], post session [49, 50], daily [40], or through weekly questionnaires [37-39]. Several studies (n=8) employed a VR orientation component [19, 37, 45-49, 51], which was beneficial to address any potential disparities in participants' level of familiarity with the technology, help subjects understand the research tasks, and allowed researchers to identify any participants experiencing cybersickness [46-49]. Winn and Jayadas [51] found that blood pressure measures were higher during the first VR exposure, indicating that participants were more nervous and less relaxed at the beginning of the study regardless of the scene content, and the authors recommended a longer VR orientation session to address this.

Simulation studies extended astronaut-like attributes to include some of the environmental and psychological conditions of space. The HI-SEAS missions replicated conditions on Mars, including mock extravehicular activities while wearing space suits, conducting habitat maintenance tasks, and implementing a 20-minute transmission delay for communication between participants and those outside the habitat [43].

Most studies utilised a single testing or interview session [20, 43-49, 51]; however, other studies used data from multiple sessions [19], over a two-week period [40, 52], or in the case of simulated missions or analogue environments, participants had access to the intervention equipment for several months [36-39, 41, 42]. Within the HI-SEAS missions, participants were instructed to use the ANSIBLE platform for 45 minutes per session, three times per week, and were otherwise free to navigate the virtual environment and create recordings to communicate with family and friends as much as they liked [38]. In Lyons and Slaughenhaupt [43], participants viewed each scene at least once and were able to revisit these at will.

**3.5 Quality Assessment Results**

No studies identified themselves as randomised controlled trial (RCT) designs. Several studies discussed blinding of participants or researchers to the experimental conditions [40, 45], randomised allocation to an experimental condition [19, 45], or randomised allocation to the sequence of exposures [20, 40, 41, 50-52], however, the randomisation processes either did not meet the MMAT criteria for an RCT design, or were not described in enough detail to determine whether the criteria were met.

As a result, all 14 of the quantitative studies were assessed against the non-randomised quantitative design criteria. Overall, the quality of these studies was high, with the majority of studies meeting 80% or more of the criteria. Where it was not determined whether the criteria were met, this was primarily due to participants not being representative of an astronaut population, despite this being the stated target population or application of the research [19, 44, 46, 47, 50]. For a few studies, the extent of incomplete measures was unclear [37-39].

There was one qualitative study [43] which met 100% of the MMAT criteria. Four studies were assessed against the mixed methods criteria. They ranged from meeting 60% of the criteria [36, 40] to 80% [50] and 100% [41]. Where it was unclear whether the criteria were met, this was due to missing values [40], lack of detail about the qualitative analysis process [36, 40], or the sample not being representative of the target population [50], and blurry figures [36].

## 3.6 Details of the Virtual Reality Interventions

All of the studies used commercially available head mounted displays, except for Finseth and Keren [19], where a high-resolution virtual reality room was used. The most reported hardware use was HTC VIVE Pro [44-48] followed by the VIVE Pro Eye [40, 51, 52], followed by Oculus Rift DK2 [20, 41]. A few studies (n=2) did not report details about hardware models used [42, 43, 50]. When contacted, the first author of both Wu and Morie [39] and Wu and Ott [38] advised that an Oculus Rift developers kit had been used for their research. Several studies (n=4) reported on the same ANSIBLE VR research within the same HI-SEAS simulation missions [36-39], so it is likely that the same hardware was used in all these studies.

There was a range of software used to present the VR scenes and content, including Unity [36, 45-48], Sketchup [50, 51], Opensimulator [38], Virtutrace [19], as well as 360° videos from feeltherelaxation.com [20], Atmosphaeres Inc [41], and a 360° artistic video called Nightsss [40]. Five studies did not specify the software used [37, 42, 43, 49, 52].

## 3.7 Useability of the Virtual Reality Interventions

Qualitative and quantitative measures were used to explore aspects of the VR user experience such as the safety of the intervention, rate of discontinuation or attrition, and likelihood of use in the future. Several studies reported cybersickness, which is a form of motion sickness in virtual environments consisting of headache, nausea, vertigo, and eye strain [53]. In Abbott and Diaz-Artiles [44], one person could not complete the protocol due to motion sickness, and two others experienced similar symptoms. In Anderson and Stankovic [41] one person reported motion sickness. In Finseth and Dorneich [46] thirteen participants withdrew due to cybersickness, and in Pochwatko and Kopec [40] the VR environment was reported to be unpleasant to move in. No participants in Winn and Jayadas [51] reported cybersickness while using VR.

In relation to presence and immersion, nature scenes were reported to be significantly more immersive than an indoor scene (Modified Reality Judgment and Presence Questionnaire, MRPJQ, $p < 0.014$) [20], and rendered scenes (MRPJQ, $p= 0.014$) [41], although preferences were highly individual [20]. Participants reported finding VR rowing scenes immersive, and differences in outcomes tended to favour the VR exercise condition over the non-VR condition [52]. Ott and Wu [37] used the System Useability Scale (SUS) to measure subjective useability of the system, and found that general ($p<.01$) and specific ($p<.01$) usability ratings increased over time (SUS, $p<.01$), whereas sense of presence declined (SFQ, $p<.01$). This may reflect an increase in participant's familiarity with, and confidence in, using the VR system, as well as reduced novelty after repeated exposures to the content.

Qualitative feedback provided in Anderson and Stankovic [41] indicated that participants were dissatisfied with some technical aspects of their VR experience including set-up times and system wiring, and that the presence of the headset and scene resolution reduced their enjoyment of the experience. Abbott and Diaz-Artiles [44] reported that one participant found the VR environment boring and repetitive when they had previously explored it, and in Pochwatko and Kopec [40] participants found interactive VR experiences more enjoyable than non-interactive experiences. When able to choose how to use the VR system, participants downloaded games to play with one another [43]. All participants in Pochwatko and Kopec [40] indicated they would participate in interactive artistic VR experiences in the future.

In Lyons and Slaughenhaupt [43], the benefits of VR included variety and novelty of sensory experiences not otherwise available in the ICE, and one participant expressed a desire for personally relevant content, such as videos of familiar or family scenes. Anderson and Stankovic [41] reported that the more isolated and confined HI-SEAS group, who lived in a small habitat with five other people for up to a year, and were unable to go outside without a space suit, found VR more beneficial than the less isolated and confined CFS-Alert group, who lived with up to 119 other people, and had access to a movie theatre, gym, games area, and other facilities. For the CFS-Alert group, interest in VR waned over time, was generally not viewed positively, and participants preferred other types of relaxation. This group expressed a desire for a greater variety of scenes containing people, animals, and action [41]. There were more alternative leisure activities available to the CFS-Alert group, whereas the HI-SEAS groups had fewer opportunities for socialisation and alternative leisure pursuits. Of the two groups, the HI-SEAS group conditions were more like the space environment.

## 4. Discussion

This scoping review sought to examine what is currently known about the use and effectiveness of VR and AI to improve mood or emotional states in people in space or other isolated and confined environments. It also aimed to identify research gaps and clarify the potential for conducting a future systematic review with meta-analysis study.

There were no eligible studies found that used AI to impact mood or emotion in space or an ICE. However, there was a diverse range of uses of VR in space and ICEs. There were two main types of uses: (i) VR as a tool to directly facilitate a change in mood or emotion, and (ii) VR as a tool for another purpose, where the impacts of mood or emotion were an additional measure of effectiveness. Studies where a change in mood or emotion was the primary purpose of the intervention, relaxation was the most popular mood or emotional variable investigated, as was the use of nature scenes to achieve this outcome. Exposure to VR nature scenes reduced negative affect using objective and subjective measures and had a high sense of presence and immersion [20]. Filmed nature scenes were more immersive and restorative than rendered scenes [42], and the addition of scents to VR nature scenes effectively reduced negative affect and anxiety following a stressful experience [44]. Biophilic designs of crew quarters, which incorporated natural elements such as wood and vegetation, were viewed more positively [50], and resulted in people feeling calmer and less nervous, than the standard, non-biophilic designs which are currently in use [51].

Although VR natural environments were successful in producing relaxation, both Lyons and Slaughenhaupt [43] and Anderson and Stankovic [41] highlighted the need for individualisation in the type of content provided to participants. Anderson and Stankovic [41] found that urban scenes also improved mood, and suggested that these scenes provided exposure to familiar city-scapes and the sense of being amongst other people, which were lacking in participants' confined environments. When participants had ongoing access to the VR equipment, they expressed a desire to use it for gaming, social interaction, and other more stimulating pursuits, rather than relaxation [43].

In Pochwatko and Kopec [40], participants described nature VR as boring, and the cinematic VR as 'terrible', and its use resulted in a decrease of positive emotions. Interactive VR experiences were found to be the most effective at alleviating the collection of negative affect symptoms (e.g., memory problems, sleep issues, fatigue, weakness) known as asthenia. Rather than relaxing, participants found the artistic VR experiences stimulating, and the authors flagged concerns about potential overstimulation in ICE simulations, where participants are constantly monitored, are in a confined space with others, and are completing demanding tasks.

Exercise is vital to counteract the detrimental physical effects of space on the human body, and using VR to help astronauts stay motivated to exercise by enhancing an exercise regime was found to be more restorative, and result in less negative affect and fatigue, than exercise without VR [52]. A VR training environment for graded exposure to reduce stress in emergencies was another application of VR technology to mood or emotion in space. This series of studies found evidence of differing responses to low, medium, and high levels of stress in VR emergency simulations set on the International Space Station [46, 47, 49], and stress responses were reduced when a milder version of the stressor was experienced prior to more intense versions of the stressor [19, 45, 48]. Stress inoculation training via VR has numerous applications to different emergency scenarios for the space environment, and would be valuable to maintain skills and emergency readiness on a long duration mission.

The ANSIBLE VR platform is a virtual world environment that was developed to support mental wellbeing and facilitate social connection in long-duration space missions [38]. Participants felt closer and more satisfied when interacting with family and friends via this VR platform [37, 39] and felt closer to humanity and that time passed more quickly using the VR platform than when using traditional, non-VR communications [38]. Interacting via this platform increased the number of positive vs negative verbalisations between crew members [36], supporting the concept that VR platforms may be useful tools for monitoring team dynamics.

Space simulations, where a group of astronaut-like participants are confined to a habitat for an extended period of time, allow us to study many of the psychological variables that impact astronauts in space. One limitation, however, is that multiple experiments may be conducted on participants at the same time. Six of the publications included in this review reported data from one or more of the HI-SEAS simulation missions which took place in Hawaii (see Table 3). Lyons and Slaughenhaupt [43] noted that for two of the HI-SEAS groups, multiple research projects involving similar VR technologies were run on the same participants concurrently, which may have affected the validity of the results through survey fatigue and recall of the interventions during later interviews.

Of the studies included in the review, nine used physiological measures such as electrodermal activity, blood pressure, pupil dilation, heart rate, and heart rate variability, in addition to self-report measures (see Table 3). These measures provide an additional objective measure of the efficacy of the interventions and are a valuable addition to subjective self-report measures. However, given the heterogeneity of applications of VR to the space and ICE environment, and the methodological differences between the studies, a systematic review with meta-analysis is unlikely to be possible at this time. Overall, virtual reality was effective in inducing change in mood or emotion both in ICE related laboratory research projects and simulations. Studies that included statistical significance (n = 16) reported at least one significant change in mood or emotion because of the VR intervention. Treatment effect sizes ranged from small (d = 0.25) to very large (d = 3.02) for graduated stress training [19, 45-49]. Many of the laboratory-based studies used a single intervention session, which indicates that VR can be an effective tool to induce change in a relatively short period of time. Cybersickness was the only negative side-effect reported, and there was no other major safety issues identified that could potentially impact the use of VR in space. Attrition was low, and primarily due to cybersickness. In combination, these findings offer a promising indication that VR produces positive outcomes for multiple purposes within the space sector, and can generally be a fast, effective, safe, and engaging tool.

Nonetheless, in addition to the limitations already identified, there were other factors which may have impacted the included studies' results. Several projects used early VR technology, which had a low resolution, required a wired connection to a computer, and has since been discontinued [20, 36-39, 41]. The wired headset interrupted participant use [52], and together with long set-up times, was noted as an implementation issue that may have reduced the effectiveness of the intervention [41]. While the technology used in the studies is likely to have been the best option available at the time, developments in design, resolution, and wireless connectivity have improved over recent years, and the user experience is likely to be much more realistic and immersive using current technology. Some studies used physiological measures of autonomic system arousal. However, the measures used in the study were primarily self-reported, which can be subject to biases such as attempts to conform to perceived social norms [54]. The lack of RCT's was also a limitation, as confounding factors such as previous diagnoses (e.g., anxiety or depression), the use of psychotropic medication, or concurrent psychotherapy (where the intervention took place over a period of time outside of a laboratory or simulation environment) were unable to be controlled for. Additionally, the laboratory-based studies were also predominantly guided by a facilitator, which may have influenced the results through factors such as demand characteristics [55].

**4.1 Future research directions**

There is currently a paucity of research into the use of AI as a psychological countermeasure in space. This is a wide gap in the knowledge base which provides many avenues for further research. There is also opportunity for further VR research to target specific mood and emotion variables, using more standardised measures and methodologies, with larger sample sizes. Research utilising higher resolution, wireless VR technologies such as the Meta Quest 3, and AI studies using the most recent large language AI models, would provide more contemporary information on potential effectiveness. New studies could also consider incorporating physiological or observational measures to corroborate self-report findings. It would be beneficial to expand upon the promising initial findings outlined in this review with studies that consider how VR can be used by participants in conjunction with different psychotherapeutic modalities and psychotropic medications, as well as by individuals experiencing the more frequent psychiatric issues found in space and other ICEs (e.g., anxiety, mood, and sleep problems [12]). A focus on the effectiveness and user experience of self-guided VR use would be a particularly useful avenue for further research, as facilitated interventions may not be available in long-duration space missions.

      For several studies in this review, participants spent limited time using VR interventions. Considering that initial interest in VR waned over time for some participants [41], future longitudinal studies investigating exposure to the intervention over a longer period of time would be helpful to identify the optimal length of interventions and assess whether impacts on mood and emotion are sustained. There was also a desire for higher stimulation, dynamic, and personalised VR experiences [43], and further research identifying the optimal range and type of content available to astronauts for a long duration spaceflight would be beneficial. A trip to Mars may take up to 3 years [24], and there are likely to be periods of monotony where astronauts' psychological wellbeing would benefit from access to novel, interesting, engaging, and regularly updated VR content as a countermeasure for boredom. The transmission of novel VR experiences via data transfer is likely to be a more realistic and achievable countermeasure for boredom and monotony during a long distance mission than the supply of physical items. VR is expected to be a valuable addition to a larger array of activities that support the psychological wellbeing of astronauts and ICE participants. Almost all the studies were from the USA, and VR and AI research with a wider range of nationalities and cultural groups would provide valuable information about how VR and AI can be best used in space by international crews. It should be noted, however, that only studies published in English were included in this review, which is likely to have limited the cultural and national range of research identified.

      Finally, none of the research included in this review used actual astronauts or was conducted in space. Further research into VR and AI should occur within this group and

setting. How the effects of cybersickness manifest in a microgravity environment, or how physical changes such as Spaceflight Associated Neuro-ocular Syndrome might impact the use of VR in space is essential health and safety information to be gathered for VR.

**4.2 Limitations of this Scoping Review**

There were several limitations to this scoping review. Firstly, studies may have been missed during the identification and screening processes, or if they were published in databases that were not included in the search. Although two reviewers completed the quality assessments, there is a degree of subjectivity in the assessments which may have biased the results. AI and VR are also rapidly developing technologies, and their applications to psychological wellbeing in the space and ICE domains are very niche, and there is understandably limited existing research in these areas to draw from. Finally, no cost analysis on the technology used in the interventions was included in the review.

5. **Conclusion**

VR has value as a psychological countermeasure in space for relaxation, social interaction, improving mood, emergency training, exercise enhancement, and testing designs. Countermeasures are likely to require multiple interventions to meet the mental health needs of the crew most effectively. The studies included in this review have established that VR has a diverse range of applications and is a valuable tool for maintaining psychological wellbeing in space and other ICE's. A lack of RCT's, small groups, and heterogeneity of VR use limit the conclusions that can be drawn. Future experimental research is required, especially examining the use of AI as a psychological countermeasure for people in space and other ICEs.

**Acknowledgements**

This research is contributing towards the Doctor of Philosophy for Jennifer Sharp.

**Data Availability Statement**

This is a review article. Data used for the findings have been reported in the manuscript itself.

**Funding**

Not applicable.

**Conflicts of interest**

There are no conflicts of interest in this project to declare.

**Author Contributions**

**Jennifer Sharp:** Conceptualization, Methodology, Investigation, Data Curation, Formal analysis, Writing - original draft. **Joshua Kelson:** Conceptualization, Methodology, Investigation, Writing – Review & Editing, Supervision. **Daryl South:** Conceptualization, Validation, Writing – Review & Editing, Supervision. **Anthony Saliba:** Conceptualization, Writing – Review & Editing, Supervision. **Muhammad Ashad Kabir:** Validation, Writing – Review & Editing.

**References**

[1]  L. Shekhtman, NASA's Artemis Base Camp on the Moon Will Need Light, Water, Elevation, Moon to Mars (2021). https://www.nasa.gov/feature/goddard/2021/nasa-s-artemis-base-camp-on-the-moon-will-need-light-water-elevation (Accessed 8 July 2023)

[2]  NASA, NASA's Moon to Mars Strategy and Objectives Development, NASA Headquarters, 300 Hidden Figures Way, SW, Washington, DC 20546, 2022. go.nasa.gov/3zzSNhp. (Accessed 8 July 2023)

[3]  T. Wolfe, The Right Stuff, Random House, 2018.

[4]  G.E. Ruff, Psychiatric problems in space flight, Dis Nerv Syst (1960) 98–101. https://spacemedicineassociation.org/download/space_medicine_classics/Psychology.pdf

[5]  D.G. Crane, Psychiatric evaluation of space flight, J Indiana State Med Assoc 55 (1962) 1623–1627.

[6]  W.F. Grether, Psychology and the space frontier, American Psychologist 17 (1962) 92. https://doi.org/10.1037/h0042950.

[7]  NASA, Human Research Roadmap, Risk of Adverse Cognitive or Behavioral Conditions and Psychiatric Disorders, 2023 https://humanresearchroadmap.nasa.gov/risks/risk.aspx?i=99, (Accessed 7th April 2024).

[8]  E. Romero, D. Francisco, The NASA human system risk mitigation process for space exploration, Acta Astronaut 175 (2020) 606–615. https://doi.org/10.1016/j.actaastro.2020.04.046.

[9] K.J. Slack, T.J. Williams, J.S. Schneiderman, A.M. Whitmire, J.J. Picano, L.B. Leveton, L.L. Schmidt, C. Shea, Risk of Adverse cognitive or behavioral conditions and psychiatric disorders: Evidence report, 2016. https://humanresearchroadmap.nasa.gov/Evidence/reports/BMED.pdf.

[10] N. Kanas, Behavioral Health and Human Interactions in Space, Springer, 2023. https://doi.org/10.1007/978-3-031-16723-2.

[11] N. Kanas, D. Manzey, Space psychology and psychiatry, Springer, 2008. https://doi.org/10.1007/978-1-4020-6770-9.

[12] E. Friedman, B. Bui, A Psychiatric Formulary for Long-Duration Spaceflight, Aerosp Med Hum Perform 88 (2017) 1024–1033. https://doi.org/10.3357/AMHP.4901.2017

[13] S.L. Bishop, From earth analogues to space: learning how to boldly go, in: D. Vakoch (Eds.) On orbit and beyond: Psychological perspectives on human spaceflight, Springer, 2012 pp.25-50. https://doi.org/10.1007/978-3-642-30583-2_2

[14] J.S. Barnett, J.P. Kring, Human performance in extreme environments: A preliminary taxonomy of shared factors, Proceedings of the Human Factors and Ergonomics Society Annual Meeting 47 (2003) 961–964. https://doi.org/10.1177/154193120304700802

[15] S. Bishop, Psychological and psychosocial health and well-being at pole station, in C.S. Cockell (Ed.), Project Boreas: A station for the Martian Geographic North Pole, British Interplanetary Society, London, (2006) pp. 160-171.

[16] A. Hamad, B. Jia, How virtual reality technology has changed our lives: an overview of the current and potential applications and limitations, Int J Environ Res Public Health 19 (2022) 11278. https://doi.org/10.3390/ijerph191811278

[17] M. Takac, J. Collett, R. Conduit, A. De Foe, Addressing virtual reality misclassification: A hardware-based qualification matrix for virtual reality technology, Clin Psychol Psychother 28 (2021) 538–556. https://doi.org/10.1002/cpp.2624

[18] V. Gushin, O. Ryumin, O. Karpova, I. Rozanov, D. Shved, A. Yusupova, Prospects for Psychological Support in Interplanetary Expeditions, Front Physiol 12 (2021) 750414. https://doi.org/10.3389/fphys.2021.750414

[19] T.T. Finseth, N. Keren, M.C. Dorneich, W.D. Franke, C.C. Anderson, M.C. Shelley, Evaluating the effectiveness of graduated stress exposure in virtual spaceflight hazard training, J Cogn Eng Decis Mak 12 (2018) 248–268. https://doi.org/10.1177/1555343418775561

[20] A.P. Anderson, M.D. Mayer, A.M. Fellows, D.R. Cowan, M.T. Hegel, J.C. Buckey, Relaxation with Immersive Natural Scenes Presented Using Virtual Reality, Aerosp Med Hum Perform 88 (2017) 520–526. https://doi.org/10.3357/AMHP.4747.2017

[21] A. Kaplan, M. Haenlein, Siri, Siri, in my hand: Who's the fairest in the land? On the interpretations, illustrations, and implications of artificial intelligence, Bus Horiz 62 (2019) 15–25. https://doi.org/10.1016/j.bushor.2018.08.004

[22] K. Zawieska, B.R. Duffy, Human-robot exploration, in: The 23rd IEEE International Symposium on Robot and Human Interactive Communication, IEEE, 2014: pp. 808–813. https://doi.org/10.1109/ROMAN.2014.6926352

[23] A.C. Strickland, Robot arrives on the space station to keep astronauts company. https://edition.cnn.com/2019/12/06/world/cimon-space-station-scn-trnd/index.html, 2020 (Accessed January, 2020).

[24] E. Linck, K.W. Crane, B.L. Zuckerman, B.A. Corbin, R.M. Myers, S.R. Williams, S.A. Carioscia, R. Garcia, B. Lal, Evaluation of a Human Mission to Mars by 2033, IDA Science and Technology Policy Institute Washington, DC, 2019. https://www.ida.org/research-and-publications/publications/all/e/ev/evaluation-of-a-human-mission-to-mars-by-2033

[25] J.R. Keebler, A.S. Dietz, A. Baker, Effects of Communication Lag in Long Duration Space Flight Missions: Potential Mitigation Strategies, Proceedings of the Human Factors and Ergonomics Society 59 (2015) 6–10. https://doi.org/10.1177/1541931215591002

[26] Z. Munn, M.D.J. Peters, C. Stern, C. Tufanaru, A. McArthur, E. Aromataris, Systematic review or scoping review? Guidance for authors when choosing between a systematic or scoping review approach, BMC Med Res Methodol 18 (2018) 143. https://doi.org/10.1186/s12874-018-0611-x

[27] L.M. Smith, The psychology and mental health of the spaceflight environment: A scoping review, Acta Astronaut 201 (2022) 496–512. https://doi.org/10.1016/j.actaastro.2022.09.054

[28] M. Gatti, R. Palumbo, A. Di Domenico, N. Mammarella, Affective health and countermeasures in long-duration space exploration, Heliyon 8 (2022). https://doi.org/10.1016/j.heliyon.2022.e09414

[29] S. Holt, Virtual reality, augmented reality and mixed reality: For astronaut mental health; and space tourism, education and outreach, Acta Astronaut 203 (2023) 436–446. https://doi.org/10.1016/j.actaastro.2022.12.016

[30]  M.D.J. Peters, C. Marnie, A.C. Tricco, D. Pollock, Z. Munn, L. Alexander, P. McInerney, C.M. Godfrey, H. Khalil, Updated methodological guidance for the conduct of scoping reviews, JBI Evidence Implementation, 19 (1) (2021) 3-10. https://doi.org/10.1097/xeb.0000000000000277

[31]  A.C. Tricco, E. Lillie, W. Zarin, K.K. O'Brien, H. Colquhoun, D. Levac, D. Moher, M.D.J. Peters, T. Horsley, L. Weeks, PRISMA extension for scoping reviews (PRISMA-ScR): checklist and explanation, Ann Intern Med 169 (2018) 467–473. https://doi.org/10.7326/M18-0850

[32]  A.I. Jabir, L. Martinengo, X. Lin, J. Torous, M. Subramaniam, L. Tudor Car, Evaluating conversational agents for mental health: scoping review of outcomes and outcome measurement instruments, J Med Internet Res 25 (2023) e44548. https://doi.org/10.2196/44548

[33]  L. Tudor Car, D.A. Dhinagaran, B.M. Kyaw, T. Kowatsch, S. Joty, Y.-L. Theng, R. Atun, Conversational agents in health care: scoping review and conceptual analysis, J Med Internet Res 22 (2020) e17158. https://doi.org/10.2196/17158

[34]  Q.N. Hong, S. Fàbregues, G. Bartlett, F. Boardman, M. Cargo, P. Dagenais, M.-P. Gagnon, F. Griffiths, B. Nicolau, A. O'Cathain, The Mixed Methods Appraisal Tool (MMAT) version 2018 for information professionals and researchers, Education for Information 34 (2018) 285–291. https://doi.org/10.3233/EFI-180221

[35]  P. Wu, J. Morie, P. Wall, E. Chance, K. Haynes, J. Ladwig, B. Bell, T. Ott, C. Miller, Maintaining psycho-social health on the way to Mars and back, Proceedings of the 2015 Virtual Reality International Conference (2015) 1–7. https://doi.org/10.1145/2806173.2806174

[36]  S.S. Galunder, J.F. Gottlieb, J. Ladwig, J. Hamell, P.K. Keller, P. Wu, A VR ecosystem for telemedicine and non-intrusive cognitive and affective assessment, in: 2018 IEEE 6th International Conference on Serious Games and Applications for Health (SeGAH), IEEE, 2018: pp. 1–6. https://doi.org/10.1109/SeGAH.2018.8401347

[37]  T T. Ott, P. Wu, J. Morie, P. Wall, J. Ladwig, E. Chance, K. Haynes, B. Bell, C. Miller, K. Binsted, ANSIBLE: A Virtual World Ecosystem for Improving Psycho-Social Well-being, Virtual, Augmented and Mixed Reality: 8th International Conference, VAMR 2016, held as part of HCI International 2016, Toronto, Canada, July 17-22, 2016. Proceedings 8 (2016) 532–543. https://doi.org/10.1007/978-3-319-39907-2_51

[38] P. Wu, T. Ott, J. Morie, ANSIBLE: social connectedness through a virtual world in an isolated Mars simulation mission, Proceedings of the 2016 Virtual Reality International Conference (2016) 1–4. https://doi.org/10.1145/2927929.2927933

[39] P. Wu, J. Morie, P. Wall, T. Ott, K. Binsted, ANSIBLE: virtual reality for behavioral health, Procedia Eng 159 (2016) 108–111. https://doi.org/10.1016/j.proeng.2016.08.132

[40] G. Pochwatko, W. Kopec, J. Swidrak, A. Jaskulska, K.H. Skorupska, B. Karpowicz, R. Masłyk, M. Grzeszczuk, S. Barnes, P. Borkiewicz, Well-being in isolation: Exploring artistic immersive virtual environments in a simulated lunar habitat to alleviate asthenia symptoms, in: 2023 IEEE International Symposium on Mixed and Augmented Reality (ISMAR), IEEE, 2023: pp. 185–194. https://doi.org/10.1109/ISMAR59233.2023.00033

[41] A. Anderson, A. Stankovic, D. Cowan, A. Fellows, J. Buckey Jr, Natural scene virtual reality as a behavioral health countermeasure in isolated, confined, and extreme environments: Three isolated, confined, extreme analog case studies, Hum Factors 65 (2023) 1266–1278. https://doi.org/10.1177/00187208221100693

[42] A. Stankovic, D. Cowan, A. Fellows, K. Binsted, J.C. Buckey, Immersive natural scenes using virtual reality for restoration in isolated confined environments, IAF/IAA Space Life Sciences Symposium 2019. https://dl.iafastro.directory/event/IAC-2019/paper/52394/

[43] K.D. Lyons, R.M. Slaughenhaupt, S.H. Mupparaju, J.S. Lim, A.A. Anderson, A.S. Stankovic, D.R. Cowan, A.M. Fellows, K.A. Binsted, J.C. Buckey, Autonomous psychological support for isolation and confinement, Aerosp Med Hum Perform 91 (2020) 876–885. https://doi.org/10.3357/AMHP.5705.2020

[44] R.W. Abbott, A. Diaz-Artiles, The impact of digital scents on behavioral health in a restorative virtual reality environment, Acta Astronaut 197 (2022) 145–153. https://doi.org/10.1016/j.actaastro.2022.05.025

[45] T. Finseth, M.C. Dorneich, N. Keren, W.D. Franke, S. Vardeman, Virtual Reality Adaptive Training for Personalized Stress Inoculation, Hum Factors 67(1) (2024) 5-20. https://doi.org/10.1177/00187208241241968

[46] T. Finseth, M.C. Dorneich, N. Keren, W.D. Franke, S. Vardeman, Designing training scenarios for stressful spaceflight emergency procedures, AIAA/IEEE 39th Digital Avionics Systems Conference (DASC), IEEE, 2020, pp. 1-10. https://doi.org/10.1109/DASC50938.2020.9256403

[47]	T. Finseth, M.C. Dorneich, N. Keren, W.D. Franke, S.B. Vardeman, Manipulating stress responses during spaceflight training with virtual stressors, Applied Sciences 12 (2022) 2289. https://doi.org/10.3390/app12052289

[48]	T. Finseth, M.C. Dorneich, N. Keren, W. Franke, S. Vardeman, J. Segal, A. Deick, E. Cavanah, K. Thompson, The effectiveness of adaptive training for stress inoculation in a simulated astronaut task, Proceedings of the Human Factors and Ergonomics Society Annual Meeting 65 (2021) 1541–1545. https://doi.org/10.1177/1071181321651241

[49]	T.T. Finseth, M.C. Dorneich, S. Vardeman, N. Keren, W.D. Franke, Real-Time Personalized Physiologically Based Stress Detection for Hazardous Operations, IEEE Access 11 (2023) 25431–25454. https://doi.org/10.1109/ACCESS.2023.3254134

[50]	A. Firth, A. Jayadas, Biophilic Design of the ISS Crew Quarters to Improve Cognitive and Physiological Health Measures, in: 2022 IEEE Aerospace Conference (AERO), IEEE, 2022: pp. 1–10. https://doi.org/10.1109/AERO53065.2022.9843279

[51]	A. Winn, A. Jayadas, T. Chandrasekera, S. Thaxton. Biophilic interventions in space habitat crew quarters to improve cognitive & physiological health, 2023 IEEE Aerospace Conference, IEEE, 2023: pp. 1–14. https://doi.org/10.1109/AERO55745.2023.10115991

[52]	N. Keller, R.S. Whittle, N. McHenry, A. Johnston, C. Duncan, L. Ploutz-Snyder, G.G.D. La Torre, M. Sheffield-Moore, G. Chamitoff, A. Diaz-Artiles, Virtual Reality "exergames": A promising countermeasure to improve motivation and restorative effects during long duration spaceflight missions, Front Physiol 13 (2022) 932425. https://doi.org/10.3389/fphys.2022.932425

[53]	J.J. LaViola Jr, A discussion of cybersickness in virtual environments, ACM Sigchi Bulletin 32 (2000) 47–56. https://doi.org/10.1145/333329.333344

[54]	S. Bauhoff, A.C. Michalos, Self-Report Bias in Estimating Cross-Sectional and Treatment Effects, in: Springer Netherlands, Dordrecht, 2014: pp. 5798–5800. https://doi.org/10.1007/978-94-007-0753-5_4046

[55]	J.F. Kihlstrom, Demand Characteristics in the Laboratory and the Clinic: Conversations and Collaborations With Subjects and Patients, Prevention & Treatment 5 (2002). https://doi.org/10.1037/1522-3736.5.1.536c